\documentclass[manuscript]{aastex}
\slugcomment{}
\shorttitle{Mid-Infrared Imaging of NGC~1068}
\shortauthors{Bock et al.}
\slugcomment{Draft version:(\today)}

\begin{document}

\title{
High Spatial Resolution Imaging of  NGC~1068 in the Mid Infrared 
%\footnotemark{1}
}
\author{J.~J.~Bock\altaffilmark{1,2}, G.~Neugebauer\altaffilmark{3}, K.~Matthews\altaffilmark{3},B.~T.~Soifer\altaffilmark{3,4}, E.~E.~Becklin\altaffilmark{5}, M.~Ressler\altaffilmark{2}, K.~Marsh\altaffilmark{6},
  M.~W.~Werner\altaffilmark{2}, E.~Egami\altaffilmark{3} and
R.~Blandford\altaffilmark{1}
}
%\footnotetext{1}{Based in part on observations obtained at the W. M. Keck Observatory which is operated jointly by the $California Institute of Technology and the University of California}
\altaffiltext{1}{Division of Physics, Math and Astronomy, California Institute of Technology,
Pasadena, CA 91125}
\altaffiltext{2}{Jet Propulsion Lab, 169-327, 4800 Oak Grove Dr., 
Pasadena, CA 91109}
\altaffiltext{3}{Palomar Observatory, California Institute of Technology,
105-24, Pasadena, CA 91125}
\altaffiltext{4}{SIRTF Science Center, California Institute of Technology
314-6, Pasadena, CA 91125}
\altaffiltext{5}{Department of Physics and Astronomy, University of
California Los Angeles, 156205 Los Angeles, CA 90095}
\altaffiltext{6}{IPAC, Jet Propulsion Lab/ California Institute of Technology,
100-22, Pasadena, CA 91125}

\email{jjb@astro.caltech.edu,
gxn@caltech.edu, kym@caltech.edu,bts@mop.caltech.edu, becklin@astro.ucla.edu, ressler@cougar.jpl.nasa.gov, kam@ipac.caltech.edu,
mww@ipac.caltech.edu, egami@mop.caltech.edu, rdb@tapir.caltech.edu}

\begin{abstract}
Mid-infrared observations of the central source of NGC~1068 have been
obtained with a spatial resolution in the deconvolved image of 0.1$''$
($\sim$7~pc).  The central source is extended by $\sim$1$''$ in the
north-south direction but appears unresolved in the east-west direction
over most of its length. About two-thirds of its flux can be ascribed
to a core structure  which is itself elongated north-south and does not
show a distinct unresolved compact source.  The source is strongly
asymmetric, extending significantly further to the north than to the
south. The morphology of the mid-infrared emission appears similar to
that of the radio jet, and has features which correlate with the images
in [OIII].  Its 12.5--24.5~$\mu$m color temperature ranges
from 215 to 260K and does not decrease smoothly with distance from the
core. Silicate absorption is strongest in the core and to the south and
is small in the north.

The core, apparently containing two-thirds of the bolometric luminosity
of the inner 4$''$ diameter area, may be explained by a thick, dusty
torus near the central AGN viewed at an angle of $\sim$65$\degr$ to its
plane. There are, however, detailed difficulties with existing models, 
especially the narrow east-west width of the thin extended mid-infrared
``tongue'' to the north of the core. We interpret the tongue as 
re-processed visual and
ultraviolet radiation that is strongly beamed and that originates in
the AGN.

\end{abstract}


\section{Introduction}

NGC~1068 is a nearby heavily obscured Seyfert~2 galaxy with a bright
active galactic  nucleus (AGN) which has been studied in nearly every
available wavelength band and at many spatial resolutions. Its
classification as a Seyfert~2 nucleus rests on the basis of narrow emission
lines. Antonucci \& Miller (1985) have, however, discovered broad,
polarized emission lines indicative of a broad line region, which
suggests the presence of a Seyfert~1 nucleus  hidden by a thick
circumnuclear torus or a heavily warped disk, a possibility we shall henceforth include in the word ``torus''. Barthel (1989) suggested that NGC~1068 might be a
prototype for many classes of AGNs. Thus NGC~1068 has provided the
framework for a unification scheme based on a dense disk of molecular gas
and dust surrounding a broad line region and Seyfert~1 nucleus; see
e.g., Krolik (1999) and references therein. In this model, Seyfert~1
and 2 galaxies are substantially similar objects viewed at different
orientations. An alternative to the torus geometry that has been
suggested by Cameron et al.\ (1993), is that the molecular material in
the nuclear vicinity is distributed in such a way that the bulk of the
gas and dust lies $\sim$100~pc or so from the nucleus.  We take the
distance to NGC~1068 as  14.4~Mpc (Tully 1988) so that 1$''$ = 72~pc;
we assume H$_o$ to be 75~km~s$^{-1}$~Mpc$^{-1}$ throughout this paper.

Since Antonucci \& Miller (1985) suggested that NGC~1068 contains an
obscuring torus, there have been numerous theoretical studies of thick
tori in AGNs.  Three that have focused on NGC~1068, mainly constrained
by the energetics, are Efstathiou, Hugh \& Young (1995), Pier \& Krolik
(1993) and Granato, Danese, \& Franceschini (1997); these give numerous
references to other theoretical models and issues.

Recently, observations of the nucleus of NGC~1068 at a range of
wavelengths with high spatial resolution have been published.
Differences exist between different images, but most find strong
nuclear emission in a roughly north-south direction; the putative torus
is presumed to lie in roughly an east-west plane. Hubble Space
Telescope (HST) observations of the inner 3$''$ of NGC~1068 in an
[OIII] filter by Evans et al.\ (1991) and Macchetto et al.\ (1994)
resolve the narrow line region into a conical shape oriented northeast
by southwest. Spectral observations by Crenshaw \& Kraemer (2000a,b)
and Kraemer \& Crenshaw (2000) of the [OIII] emitting gas have further
delineated the cone. Speckle observations at the 2.2~$\mu$m
diffraction limit of the Keck 10-meter Telescope show that about half
of the nuclear flux at 2.2~$\mu$m within a diameter of $\sim$1$''$
comes from an extended region, essentially north-south but orientated slightly
west of north, 0.3$''$ (22~pc) in length on either side of a point
source which is less than 0.03$''$ (2~pc) in size (Weinberger et al.\
1999).  Rouan et al.\  (1998) have used adaptive optics to obtain
images at 1.2, 1.6 and 2.2 $\mu$m with higher dynamic range  but poorer
spatial resolution than Weinberger et al. and see evidence of elongated
structure at a position angle of P.A.~$\approx$~102$\degr$ which might
be tracing the putative torus. Marco \& Alloin (2000), observing at 3.5
and 4.8~$\mu$m,  see an unresolved core with a full width at half
maximum (FWHM) $<$~0.2$''$ (16~pc), a north-south elongation about
1$''$ (70~pc) in length, plus a ``disk-like'' structure with a diameter
$\sim$1.0$''$ (70~pc) at a P.A.~$\sim$~100$\degr$ which they interpret
as the dusty torus invoked in unification schemes.

Gallimore et al.\ (1996a) and Gallimore, Baum, \& O'Dea, (1996c) have
summarized the sub-arcsecond radio structure of NGC~1068 as consisting
of three components in a northeast by southwest line. A 10$''$
(kiloparsec) scale radio jet joins the southern and central radio
components and traces emission to sub-arcsecond (tens of pc) scales.
The southern-most radio peak consists of two clumps, the brighter of
which has a relatively flat radio spectrum and  is associated with
H$_2$O and OH masers characterizing warm and dense molecular gas and
showing Keplerian motion. Gallimore et al.\ (1996c) consider this
region (called S1) as most likely being at the location of the central
engine of NGC~1068 because of these features.  Gallimore, Baum \& O'Dea
(1997) have imaged S1 at 8.4~GHz and give the first direct image of a
pc-scale ionized gas disk surrounding the AGN in NGC~1068 when they
trace a set of compact sources of overall extent 0.15$''$ at a position
angle of $\sim$110\degr. S1 appears to also coincide with the apex of
the [OIII] cone within the astrometric uncertainties between the visual
and radio images (0.2$''$; Evans et al.\ 1991).    At millimeter
wavelengths, Schinnerer et al.\ (2000) see two emission knots in the
$^{12}$CO(2-1) line about 1$''$ east and west of the nucleus surrounded
by a ring of emission with radius $\sim$1.5$''$ (100~pc).

The bolometric luminosity of NGC~1068 is determined mainly in the mid
infrared where it is extremely bright; its flux distribution
peaks at $\sim$25~$\mu$m. If the infrared emission is isotropic, the
bolometric luminosity of NGC~1068 is
$\sim$2.5--3.0$\times$10$^{11}~$L$_\sun$, most of which is emitted
between 10 and 100 $\mu$m. This includes, however,  the emission from
the ``3~kpc ring'',  a disk about 30$''$ in diameter, which was
observed at 10, 60 and 158~$\mu$m by Telesco et al.\ (1984) and found
to account for almost half the luminosity of NGC~1068.  Telesco et al.
attributed this emission to a molecular cloud with ongoing star
formation.  Subsequent radio measurements by Planesas, Scoville \&
Myers (1991) and Papadopoulos, Seaquist \& Scoville (1997) further
delineated the properties of the molecular cloud.  Finally, Laurent et
al.\ (2000) obtained CVF spectra between 5 and 16~$\mu$m which allowed
them to clearly separate the mid-infrared emission from the nucleus and
from the outer ring. Based on the observations detailed below we will 
take the bolometric luminosity of the central area 4$''$ in diameter 
of NGC~1068 to be 1.5$\times$10$^{11}~$L$_\sun$. 

Hard  X-ray observations of NGC~1068 provide another estimate of the
intrinsic AGN luminosity.  A strong Fe K line, first detected with
Ginga by Koyama et al. (1989), indicates that the observed x rays are
scattered and reflected emission, and that the direct line of sight to
the AGN is completely blocked by a Compton-thick material, a result
confirmed by observations with ASCA (Ueno et al.\ 1994) and BeppoSax
(Matt et al.\ 1997). The intrinsic 2--10~keV luminosity of NGC~1068 is
estimated to be on the order of $\sim$1.1$\times$10$^{10}$L$_\sun$
(Iwasawa et al.\ 1997; rescaled to H$_o$ = 75~km~s$^{-1}$~Mpc$^{-1}$).  
If L$_{bol}\sim30$L$_{\rm2-10~keV}$ (Matt et
al. 2000), the intrinsic AGN bolometric luminosity is estimated to be
$3\times10^{11}$L$_{\sun}$. 

The reduced obscuration at mid-infrared wavelengths allows a detailed
examination of the nuclear structure, and means that mid-infrared
observations are crucial in determining the morphology of the nucleus.
The previous mid-infrared observations of NGC~1068 with the  highest
spatial resolution are those of Bock et al.\  (1998) who made
diffraction limited observations (FWHM $\sim$0.5$''$)  with the Hale
200-inch Telescope at 8.8, 10.3 and 12.5~$\mu$m; see Bock et al.\  for
references to previous  mid-infrared work.   Bock et al.\  report a
central peak with structures extending about 1$''$ north-south which is
unresolved east-west and coincides with one wall of the ionization cone
seen by Evans et al.\ (1991).  In this paper we report similar
observations but made at wavelengths from 7.9~$\mu$m to 24.5~$\mu$m
with the 10-meter Keck Telescope and accordingly having twice the raw
spatial resolution as the Palomar observations. The present
observations were limited to the inner 4$''$ of the galaxy and pertain
to this extreme nuclear region only.

\section{Observations}

Observations of NGC~1068 were conducted from the Keck~2 10-meter
Telescope on the nights of 1~October and 3~October~1998 with the
mid-infrared camera MIRLIN (Ressler et al.\  1994).  MIRLIN is based on
the HF-16 128~$\times$~128 Si:As BIB focal plane array with a plate
scale at the Keck Telescope of 0.138$''$~pixel$^{-1}$, resulting in
Nyquist-sampling of the diffraction-limited Airy function for
$\lambda>$12~$\mu$m.  In order to characterize and monitor the point
spread function and to obtain photometric calibration, observations of
NGC~1068 were interleaved with those of a bright infrared stellar
source, either BS=HR~0911 ($\alpha$~Cetus) or BS=HR~1457
($\alpha$~Taurus).  The sequence of observations shown in Table~1 was
chosen so that the infrared standards were observed at a similar
elevation angle as NGC~1068.  The targets were observed in four
mid-infrared spectral bands, centered on 7.91~$\mu$m
($\Delta\lambda$=0.76~$\mu$m), 10.27~$\mu$m
($\Delta\lambda$=1.01~$\mu$m), 12.49~$\mu$m
($\Delta\lambda$=1.16~$\mu$m), and 24.48~$\mu$m
($\Delta\lambda$=0.76~$\mu$m).

In order to accurately subtract thermal emission from the atmosphere
and telescope, each target was observed with a standard ``chop and
nod'' strategy.  A chopped image was created by coadding exposures in
the on and off beam of the chopping secondary mirror.  The telescope
was then nodded to remove any offset induced by the chopper, and the
observation repeated to create the ``chop-nodded" image.  Typical
integration times were 80 to 100~ms in each beam of the chopper, with a
total of 4 - 5~s of integration on the sky spent on each of the four
images. Table~1 is a log of the observations.  Each observation listed
in Table~1 consisted of a sequence of at least three chop-nodded images
at each wavelength, dithered by $\sim$~0.7$''$ between images in order
to sample different pixels on the array.  The telescope was guided
during the observations on nearby stars using an automated optical
guider, always selecting the guide star so as to avoid diffraction
spikes from the target.  The chop and nod amplitudes were each 5$''$ so
as to keep the image of the target on the focal plane array for maximum
observing efficiency.  To avoid any smearing of the images during the
exposures, the settling time during each chopper cycle was 15~ms and
the settling time between nod positions was 15~s. In order to confirm
the PSF was not elongated due to chopping or nodding, the chopping was
north-south and nodding east-west on 1~October~1998 and the chopping
was east-west and nodding north-south on 3~October~1998.

\section{Data Reduction and Analysis} 
\subsection{Data Reduction} 
The raw data of NGC~1068 and the PSF calibration stars were combined by
dividing each exposure by a flat field frame and interpolating over bad
pixels from a bad pixel mask.  The four sub-images in each
chopped-nodded frame were combined, spatially filtered to reduce noise
at high spatial frequencies, and rotated to a reference angle.  These
dithered sub-images were then coadded to produce a single image for
each image number in Table~1.  Because the telescope operates on an
altazimuth mount, the characteristic diffraction spikes induced by the
segmented primary and secondary spiders appear to rotate in equatorial
coordinates during the night.  In the mid infrared, the relative
intensity of the diffraction spikes is small, roughly 1\% of the peak
intensity of the PSF.  In the case of NGC~1068,  each image was rotated
to equatorial north.  For the PSF images, each raw image was either
rotated to equatorial north, for purposes of creating the deconvolved 
images of NGC~1068, or, for purposes of creating the deconvolved 
images of the PSF,  rotated to a common elevation angle thus keeping 
the orientation of the diffraction spikes fixed.

\subsection{Image Deconvolution}

The images were rebinned to a finer pixel scale
(0.034$''$~pixel$^{-1}$) and deconvolved using a Richardson-Lucy
maximum likelihood solution subject to a positivity constraint as
developed by Richardson\ (1972) and Lucy\ (1974).  The deconvolved
images of NGC~1068 and the PSF are shown in Figure~1. As the
deconvolved image depends critically on the stability of the PSF
image, the matching of the PSF calibrator in elevation angle and the
interleaved observations of NGC~1068 and the PSF should give optimal
results.

To estimate the stability of the PSF image,  one image was formed from
the normalized sum of the first, third and fifth images of the PSF from
Table~1 and a second image was formed from the normalized sum of the
second and fourth images from Table~1.  This linear combination of PSF
images results in a small effective difference in elevation angle
between the two combined images. The deconvolution of the first of
these combined PSF images by the second results in the compact images
of the PSF shown in the third column of Figure~1 (FWHM $\sim$0.1$''$).
The combined image of HR~0911 obtained on 3 October was also
deconvolved by the combined image of HR~0911 obtained on 1 October,
with an excellent result and negligible residual features.  A more
detailed discussion of various tests made to verify the deconvolution
procedure is given in the Appendix.

MIRLIN only provides sub-critical spatial sampling of the PSF for
wavelengths $\lambda\leqq$ 12~$\mu$m.  Because the image deconvolution
requires critically sampled images, all of the unprocessed images taken
at 7.9 and 10.3~$\mu$m were first smoothed by convolving them with a
2-D Gaussian function of FWHM 0.2$''$ before proceeding with the data
reduction.  This smoothing has the effect of degrading the intrinsic
spatial resolution at 7.9 and 10.3~$\mu$m  to the intrinsic spatial
resolution at 12.5~$\mu$m.

\subsection{Photometry}
Photometric measurements were made of the raw images before any
deconvolution algorithms were applied. Photometry was based on the
measurements  of HR~1457  and  HR~0911 whose magnitudes were determined
by a combination of an inter-comparison of several measurements taken
on several nights, by an extrapolation based on their effective
temperature and by a consistency with the IRAS 12 and 25 $\mu$m
measurements. The magnitudes adopted at wavelengths  7.9, 10.3, 12.5
and 24.5~$\mu$m are  -2.99, -3.04, -3.07 and -3.03~mag for HR~1457 and
-1.86, -1.89, -1.92 and -1.96~mag for HR~0911. The ``sky'' was taken as
an annulus 4.6 to 6.0$''$ in diameter. The conversion from magnitude
to flux density followed the prescription given in the Explanatory
Supplement to the IRAS Catalog (Beichman et al.\ 1985); i.e., 0.0~mag =
63.2, 38.0, 26.2 and 7.0~Jy at 7.9, 10.3, 12.5 and  24.5~$\mu$m.

\section{Results}
\subsection{Imaging}

The combined images of NGC~1068 and of the PSF calibration stars are
given as the ``raw images'' displayed in Figure~1.  The PSF was
diffraction limited at 12.5 and 24.5 $\mu$m, varying from a FWHM of
0.3$''$ at 12.5~$\mu$m to 0.6$''$ at 24.5~$\mu$m, with the Airy
diffraction rings apparent in each of the raw images.  At 7.9 and
10.3~$\mu$m the FWHM was $\sim$0.4$''$ due primarily to the smoothing
described in Section~3.2. The ratio of the peak flux density per pixel
to the noise level in the raw images varies from $\sim$300 in the
24.5~$\mu$m image to $\gtrsim$1,000  in the other  three images.

The four images were deconvolved as described above and the four
deconvolved images of NGC~1068 and of the PSFs, derived as
described above, are given in Figure~1. The FWHM of the
deconvolved PSF in the 12.5~$\mu$m image is 0.05$''$; that at
7.9, 10.3 and 24.5~$\mu$m is 0.07, 0.10 and 0.12$''$. The maximum
contour shown is of pixels with a flux density surface brightness
of 0.9$\times$[maximum flux density surface brightness per
pixel]. The minimum contour shown is down by a factor of 140 from
the peak surface brightness in the 7.9, 10.3, and 12.5~$\mu$m
images, and by a factor of 70 in the 24.5~$\mu$m image. The
uncertainties increase towards the center, to roughly 5\% of the
peak surface brightness at the peak in the case of the
12.5~$\mu$m image. Based on the reproducibility of the
deconvolved images described in the appendices, the confidence in
the reproducibility in the 12.5~$\mu$m image extends to  the
minimum contour shown and is somewhat less at the other
wavelengths.  In the 7.9~$\mu$m image the feature to the
southwest is possibly spurious, as it does not reproduce in the
first and second deconvolved images (see Figure~8).  A full
accounting of the sources of uncertainty is given in the
appendices. Figure~8 demonstrates the reproducibility of the
deconvolved images from independent data sets. 

As seen in Table~1, the image taken at 12.5~$\mu$m had the 
largest number of multiple redundant observations.  It has the 
highest angular resolution and highest signal
to noise ratio, i.e., dynamic range, of the four images obtained.
For these reasons we have chosen to duplicate the deconvolved
12.5~$\mu$m image in Figure~1b. We want to emphasize, however,
that all four images, in particular that at 24.5~$\mu$m, show
qualitatively the same features. 

Each of the images is dominated by a central peak which contains
$\sim$~one third to one-half of the flux in a 4$''$ diameter beam
depending on the limits ascribed to the peak, which are ambiguous. The
FWHM of the peak at 12.5~$\mu$m is 0.26$''$; the widths at the other
wavelengths are comparable. The peak contains extended emission, and,
as shown below, is significantly more extended than can be accounted
for by a point-like source.  The emission above $\sim$10\% of the peak
surface brightness is generally extended in a well resolved north-south
line (position angle measured east from north $\sim$10$\degr$) about
1$''$ (70~pc) in length. Both the central peak and this extended
emission are compact in the east-west direction.  With the present
deconvolution, they are formally slightly larger than the PSFs in the
east-west direction; the measurements imply an intrinsic width of the
peak at different wavelengths between $\sim$0.05$''$ and $\sim$0.1$''$.
This is sufficiently small that we do not claim that we have resolved
the peak in the east-west direction.  At the different wavelengths, the
peak blends smoothly into the clearly extended emission in the
north-south direction.  Based on a comparison of the raw images and
re-convolved images, we estimate that for distances $\gtrapprox$0.5$''$
from the peak, the average uncertainty per pixel in the deconvolved
images ranges from $\sim$4\% of the peak value per pixel at 24.5~$\mu$m
to $\sim$1\% in the other three images.  The uncertainties generally
increase towards the peak; see the Appendix for details.

The fainter extended emission follows a similar pattern in the four
wavelengths bands. The northern extension appears to first bend
slightly west, $\sim$0.35$''$ north of the infrared peak, and then,
$\sim$0.5$''$ north of the infrared peak, becomes more extended and
bends east by $\sim$60$\degr$.   The southern emission, which is less
extended than the northern emission, also becomes more diffuse south of
the largest infrared peak. A faint source appears $\sim$0.75$''$ to 
the northeast of the main infrared peak.

The deconvolved images in Figure~1 are qualitatively like  the
deconvolved images at $\sim$0.2$''$ deconvolved angular resolution 
from the Hale 5-m Telescope described by Bock et al.\ (1998).   Both show
a linear structure about 1$''$ in the north-south direction that is
unresolved in the east-west direction. The bend to the east present in
the lower contours in Figure~1, however, is not apparent in the lower
resolution images. Both the Keck and Palomar images are in general
agreement with the deconvolved mid-infrared images from Braatz et
al.\ (1993) with much lower angular resolution (deconvolved FWHM =
0.7$''$ at 12.4~$\mu$m).  Braatz et al. show an image which is oriented
basically northeast by southwest, but their resolution  is too low to
distinguish shapes within the extension. The northeast orientation in
their image may reflect a low resolution observation of the  bend seen
at Keck. The images are also in qualitative agreement with the lower
resolution images of Cameron et al.\ (1994) although the northeast
elongation observed by these authors is resolved into at least two
essentially linear features.

\subsection{Photometry }

The magnitudes  of NGC~1068 within an artificial 4$''$ diameter beam
are given in Table~2. The uncertainties listed  include both
statistical and estimated calibration uncertainties.  The flux
densities reported here are  similar to  those reported in the 
literature. In view of the well established dependence on beam 
size plus the known near-infrared variability in the luminosity 
of the nucleus of NGC~1068,  which is as
large as a factor of two in two decades at 3~$\mu$m (see, e.g., Glass
1997) any small difference is not disturbing.

Photometric magnitudes of NGC~1068 were also obtained in a beam
with a 2$''$ diameter.  The average differences between the 2$''$ and
4$''$ diameter beams, which are not subject to many of the systematic 
uncertainties of the photometry, are also given in Table 2.

\section{Discussion}
\subsection{Overview} 

An overview of the energetics of NGC~1068 and how the present
observations fit into the luminosity of NGC~1068 is provided by
Figure~2 which compares the flux per octave of the Keck observations
with the overall spectral energy distribution measured from NGC~1068.
The latter has been determined from the published literature and
reflects large beam measurements such as the IRAS measurements.
Figure~2 establishes that the wavelengths used in the present
observations sample the bulk of the emission from NGC~1068.

The arrows in Figure~2 indicate upper limits on the flux per octave
measured in a Gaussian of FWHM equaling the east-west width of
NGC~1068. In the four wavelength bands they represent 35, 14, 21 and
21\% of the total, and represent estimates of the emission which can be
ascribed to an unresolved point source.  The estimate is an upper limit
since the width of NGC~1068 is certainly greater than the FWHM of the
deconvolved PSF.  We are forced to the conclusion that only a small
fraction of the observed luminosity in NGC~1068 comes from a point like
nucleus; even in  close to the central peak there is a significant
component of extended emission.

Emission from the low surface brightness disk $\sim$35$''$ in diameter
corresponding to a 3~kpc diameter ring of molecular clouds in which
star formation is occurring, is included in the overall fluxes in
Figure~2 (Telesco et al.\ 1984). Telesco et al.  conclude that about
half the bolometric luminosity of NGC~1068 comes from  the 3~kpc ring
and  estimate that  at 10~$\mu$m $\sim$20\% of the observed flux
density comes  from the disk.    The disk is much colder than the
nuclear source; Telesco et al. estimate that the 8 -- 25~$\mu$m
radiation from the disk is only about 20\% of the total.  In contrast,
at 100~$\mu$m, most of the emission is from the disk and only a small
fraction is from the nucleus. From the Keck data and Figure~2 we
estimate that $\sim$two-thirds of the bolometric luminosity comes from
the central area with 4$''$ diameter. I.e., the 3~kpc ring contributes
only $\sim$one-thirds the bolometric luminosity. This apparent
discrepancy between the Keck data and the data  of Telesco et al. can
be attributed to the uncertainties in the models  and in  the
calibrations, including the wide bandwidths, plus the inherent
difficulties of observations in the far infrared. It is best resolved
with high spatial resolution observations at 60 and 100~$\mu$m such as
will be available with the SOFIA mission. Since it does not affect the
following discussion, it will not be pursued further.

\subsection{Comparison with [OIII] and Radio Emission} 

Figure~3 shows the 12.5~$\mu$m contour map from the Keck observations
superimposed on the HST [OIII] image of Macchetto et al.\ (1994).
Since no absolute astrometry was obtained for the Keck images, the
juxtaposition of the infrared central peak with the apex is uncertain.
The  mid-infrared peak was located at the scattering center as determined by
Kishimoto (1999) who has recently carried out a reanalysis of the
polarization data (Capetti et al.\ 1997) to determine the position of
the nucleus of NGC~1068 as defined by ultraviolet  polarimetry.
Independent determinations of the position of the nucleus from
astrometry at near-infrared wavelengths (Thatte et al.\ 1997) and
mid-infrared wavelengths (Braatz et al.\ 1993)  agree within the quoted
uncertainties. With the choice of Kishimoto's central position, 
the line of the
12.5~$\mu$m  emission lies just inside the  ionization cone, adjacent
to one `` wall''. Also, this choice of alignment gives a good
correlation between the mid-infrared emission and [OIII] clouds B-F,
and thus provides independent confidence in the efficacy of the 
adopted deconvolution.

The 12.5~$\mu$m contour is superimposed on the 5 GHz map of Gallimore
et al.\ (1996c) in Figure~4. Again, since no absolute astrometry was
obtained for the Keck images, the juxtaposition of the infrared central
peak with the radio features is uncertain. A shift of 0.3$''$ would
more completely overlay the mid-infrared features with the radio jet
and cannot be ruled out. The 12.5~$\mu$m central peak has been located
on S1, the brighter of the southern components contained in the radio
nucleus and the source Gallimore et al.identify with the center of
NGC~1068. It has a flatter radio spectrum than the other radio sources
and is the location of the torus observed by Gallimore et al.\ (1997).
The qualitative agreement of the central portions is striking. Both
images show a bend to the east, although the infrared bend is further
to the north than the radio bend.

\subsection{Comparison with 3.5 and 4.8~$\mu$m Observations}

At the brighter levels, the appearance of the  present mid-infrared
images agrees qualitatively with the near-infrared images of Marco \&
Alloin (2000) in showing extended emission along the north-south
direction, the preferential direction of the axis of the radio emission
and of the ionizing cone. As in the mid infrared, the central core 
is unresolved in the near-infrared images (FWHM $\leqq$0.12$''$). 

The agreement at the lower intensity contours is not so good.  At both
3.5 and 4.8~$\mu$m, Marco \& Alloin (2000) see evidence both
$\sim$0.5$''$ east and west of the central peak for a putative torus at
an intensity level roughly 1\% of the peak surface brightness
observed.  The flux density at 4.8~$\mu$m of the clumps defining this
torus is $\sim$0.15~Jy.  No sign of this torus is seen in the present
mid-infrared observations. If the radiation is thermal, at a reasonable
temperature, it should also be present in the  7.9~$\mu$m deconvolved
image. Tests made by placing an artificial point source 0.5$''$ from
the central peak of NGC~1068 indicate that at 7.9~$\mu$m a source with
a flux density of $\sim$0.1~Jy or 0.5\% of the total observed flux
density at 7.9~$\mu$m  could easily be seen; see the Appendix.
Calculations using the thermal parameters of Draine \& Lee (1984) show
that  ordinary astronomical graphite grains emitting 0.1~Jy at
7.9~$\mu$m would have to be hotter than 600~K to emit 0.15~Jy at
4.8~$\mu$m. Ordinary astronomical graphite grains re-radiating thermal
radiation after being heated by a source with a luminosity of
1.5$\times$10$^{11}$L$_\sun$ and 35~pc distant, would come to a
temperature $\sim$~245~K. Thus, the grains cannot be heated purely by
the central source unless the emissivities are vastly different from
commonly accepted values, or the lumps derive their energy from an
entirely different source.

\subsection{Silicate Absorption}
Kleinmann, Gillett \& Wright (1976) obtained a spectrum of NGC~1068
from 8 to 13~$\mu$m with $\sim$2\% spectral resolution in a 5$''$
diameter beam which demonstrated that NGC~1068 has a pronounced
absorption feature centered at $\sim$9.7~$\mu$m attributed to
silicates. More recent ISO-SWS spectra by Sturm et al.\ (2000) have
delineated a broad absorption feature extending from $\sim$8.6~$\mu$m
to $\sim$10.0~$\mu$m which is centered at 9.4~$\mu$m rather than
9.7~$\mu$m, but the spectrum is probably contaminated because of the
large ISO beam. The wavelength range of the 10.3~$\mu$m band used in
the present observations overlaps about half of this absorption and
thus provides a measure of any strong silicate absorption morphology.
Figure~5 gives the spectral energy distribution at the five locations
shown. These locations were chosen to approximate local maxima in the
12.5~$\mu$m image after the four deconvolved images of Figure~1 were
smoothed to $\sim$0.2$''$ FWHM.  It should be noted that, because the
images at all four wavelengths were smoothed to the lowest resolution,
this is a robust result.

The figure shows strong absorption on the peak and just to the south.
There is a smooth  continuum, perhaps even silicate emission, to the
north and northeast. The most straight forward explanation is that we
are seeing an attenuating screen of silicate  which is  more
concentrated over the central peak of mid-infrared emission than in
other regions. The absence of a silicate absorption feature in certain
regions of the image could, of course, be due to purely geometrical
effects such as the lack of temperature gradients along the line of
sight.

\subsection{Temperatures}

In order to estimate the brightness temperature of the central peak,
the source size was taken as that area which contained pixels in the
deconvolved image with a  flux density per pixel greater than half the
maximum value. For the 12.5~$\mu$m image, that with the best
resolution, the size of this area was $\sim$0.4~{\sq}pc
(0.006~{\sq}$''$) and the  resulting brightness temperature was 193~K.
Brightness temperatures for the other wavelengths, similarly defined,
ranged from from 118~K at 24.5~$\mu$m, with a source size
$\sim$2.7~{\sq}pc, to 205~K at 7.9~$\mu$m, with a source size
$\sim$1.8~{\sq}pc.

A 12.5 to 24.5~$\mu$m color temperature was calculated from the
deconvolved images at these wavelengths after first smoothing the data
to a common FWHM of 0.2$''$.  Only flux densities per pixel larger than
three times the estimated noise per pixel, i.e.  0.03$\times$[maximum
flux density/pixel] at 12.5 $\mu$m and 0.12$\times$[maximum flux
density/pixel] at 24.5 $\mu$m were used. The resultant color
temperatures, shown in Figure~6, are quite sensitive to the alignment
of the two images, but in the central portion are $\sim$265~K. The
color temperature is highest in the central peak at 269~K and,
significantly, there is no smooth dropping of the temperature along the
extended north-south emission. In fact, the color temperature north of
the central peak falls to $\sim$~216~K $\sim$0.25$''$ (20~pc) north of
the peak and then rises again to $\sim$263~K $\sim$0.4$''$ (30~pc)
north of the peak.

Calculations using the averaged emissivity parameters of Draine \& Lee
(1984) indicate that an optically thin cloud of pure silicate grains in
thermal equilibrium with a central source of
1.5$\times$10$^{11}$L$_\sun$ would have  12.5 to 24.5~$\mu$m color
temperatures of 270~K and 220~K at distances of 20 and 30~pc.  If
graphite grains were assumed, the color temperatures the same distances
from the source would be 320~K and 270~K. The physical temperatures are
predicted to be about 10 to 20~K cooler. Thus the measured color
temperatures are in the range as expected for heating by a central
source, but do not show a smooth behavior with radial distance expected
for a uniform distribution of dust grains in such a situation.

An obvious possibility is that the warm emission is from single photon
heated small grains. This has, for example, been invoked in the
starburst galaxy M82 to explain an increase in the color temperature
with distance from the center (Telesco et al.\ 1991).  Although this
possibility cannot be excluded, we consider it unlikely since if it
were the explanation for the warm emission, it would also be necessary
to explain why, in Figure~5,  there is  no sign of the PAH emission
which is ubiquitous with single photon heating.

Of course, beaming of a central source could increase the predicted
temperature.  An explanation for the results of Antonucci \& Miller
(1985) could invoke beaming of the central source. Baldwin, Wilson \&
Whittle (1987) have shown that the intensity of the central source
could be as much as 200 times greater along the radio axis than to the
Earth. If \textbf{f} is the ``beaming factor'' by which the luminosity
of the central source in the direction of the dust cloud exceeds that
towards Earth, the resultant temperature would be increased by a factor
of \textbf{f}$^{1/4}$. Thus temperatures as high as 500~K  could be
obtained with beaming factors in the hundreds, although we  do not
consider such extreme beaming necessary.

\subsection{Interpretation} 

The most prominent feature of the present observations is the highly
asymmetric---almost linear---nature of the images, the well resolved
extent north-south, and the extreme narrowness east-west. A dominant
feature of this emission is a relatively constant 12.5 to 24.5~$\mu$m
flux ratio or color temperature. We will designate  the narrow lobe to
the north  of the dashed line in Figure~1b as the ``tongue''. Also
prominent is a ``core'', a name we assign to the region south of the
dashed line in Figure~1b. We acknowledge that the dividing line between the two areas is arbitrary and is partially based on the model
described below. The core contains about two-thirds the flux per octave
in the mid infrared which we identify with the bolometric luminosity of
the central source within an area 4$''$ in diameter, while the tongue
contains about one-third. The core is aligned so its north-pointing
axis points in a direction slightly west of north while the emission in
the tongue curves slightly east of north.

As a corollary to the feature noted above, we emphasize that the
mid-infrared observations, by their small extent, convincingly
associate a significant fraction, more than half, of the bolometric
luminosity of NGC~1068 with an AGN rather than with star bursts.  Our
interpretation of  the images is divided into a discussion of
the core and a discussion of the tongue.

\subsubsection{The Core}

The core region is itself asymmetrical; the brightness of the emission
in the north of the core is significantly greater than that to the
south. The emission coming from the core can be understood if  there is
a  very dense torus originating  within a parsec or so of the AGN.
Granato et al.\ (1997) have proposed a simple model consisting of a
torus extending from $\sim$0.2~pc to tens of pc. (Note, however, that
this assumes that the torus has uniform density whereas the HCN
observations of e.g. Tacconi et al.\ (1994) suggest that the gas
observed at $\sim$100pc is quite inhomogeneous.) The asymmetric
structure of the core image is a natural consequence of the inclination
of the plane of the  torus to the line of sight. We are seeing emission
from the heated inside edge to the north, while the heated southern
side is partly obscured by the torus itself. Granato et al. invoke an
observation angle of 65$\degr$ and predict an asymmetry of $\sim$14:1
intensity ratio between the flux in a 0.2$''$ diameter beam at the peak
of the core and one in the lobe to the south. The observations yield a
ratio  of $\sim$10:1 for the flux density within a 0.2$''$ diameter
beam at the core and at the lobe to the south (positions b and a in
Figure~5). The predictions of Granato et al. give a separation of
0.3$''$ between the core peak and the southern lobe, in good agreement
with the  measured separation of 0.4$''$.

Although the observations and models for the core region are in
excellent agreement in some aspects, in others they differ. The models are clearly not attuned to the small scale structure observed. The
east-west width of the radiation pattern in the model of  Granato et
al.\ (1997) is significantly larger than observed. The modeled
10~$\mu$m surface brightness has an east-west FWHM $\sim$0.15$''$ while
the deconvolved 12.5~$\mu$m observations are barely, if at all,
resolved east-west with an east-west FWHM $\lesssim $0.05$''$.  The
modeled 10~$\mu$m surface brightness drops to 10\% at a full width of
0.4$''$, while in the deconvolved 12.5~$\mu$m image the measured full
width at 10\% is closer to 0.1$''$. The model clearly requires
significantly deeper silicate absorption in the spectral energy
distribution of the southern component compared to that of the northern
source in the core. Figure~5 shows silicate absorption at both the
southern locations (positions a and b), but whether the extreme
southern source displays the increased absorption required to attenuate
the signal from the near side of the torus is uncertain, depending  on
the continuum temperature at the two locations. The high aspect ratio
of the mid-infrared images, and whether the southern lobe we identify
as part of the core is indeed associated with a torus remain challenges
for torus models to address.

\subsubsection{The Tongue}

The tongue is a powerful mid-infrared component in NGC~1068 associated
with the western boundary of both the [OIII] emission and the radio
continuum emission (Evans et al.\ 1991; Macchetto et al.\ 1994;
Gallimore et al.\ 1996c).  On spectral grounds we identify this with
re-emission  by dust of visual and ultraviolet radiation concentrated 
in an ionization cone which most likely originates in the nucleus of NGC~1068. 

At this stage we note and comment upon three explanations for 
the formation of the ionization cone in NGC~1068.
\begin{description}
\item[Relativistic beaming] The ultraviolet continuum could be 
produced nonthermally by a pair of relativistic jets and beamed 
along the jet direction through relativistic aberration.  There 
is no evidence, however, that the jets associated with Seyfert 
galaxies, in contrast to those associated with spiral galaxies, 
are relativistic and we consider that this explanation is unlikely.

\item [Dust absorption] The ultraviolet continuum may be produced 
in a roughly isotropic source, like an accretion disk, and cool 
dusty gas at low latitudes -- the torus -- then remove most of 
it only permitting escape along the two polar directions. As the 
solid angle subtended by the ionization cone is $\sim$0.1 of the 
sky, we would expect the power radiated by the torus to be at 
least ten times that associated with the ionization cone. We 
have already shown, however, that the power re-radiated at 
mid-infrared wavelengths from the northern ionization cone is one 
third that from the core, and this is a lower bound on the 
luminosity flowing along the ionization cone. As the core luminosity 
is dominated by the mid-infrared power reported here, we conclude 
that the core is under-luminous by at least an order of magnitude 
for dust absorption to account for the formation of the cones. (See 
Storchi-Bergmann, Mulchaey, \&  Wilson\ (1992) for a related argument.)

\item [Electron scattering] Free electrons have a very high albedo 
at ultraviolet wavelengths, and a hot, ionized torus, probably 
orbiting inside a cool dusty torus will scatter most of the 
radiation from an isotropic ultraviolet source into the polar 
directions.  This is our favored explanation for the ionization 
cone. (See Miller, Goodrich \& Matthews, 1991.)  Note that the 
x-ray observations  also suggest the presence of a hot torus.
\end{description}

The above picture is strengthened by the observations of Capetti
et al.\ (1995)  who measured wavelength-dependent, polarized visual
and  near ultraviolet continuum emission from the ionization cone which
appears to be concentrated at the edges of the radio emission (Capetti
et al.\ 1997), just as in  the mid-infrared observations  reported
here. This suggests that we are under-estimating the luminosity of the ionization cone.

In view of the energetic considerations given above, it is appropriate
to compare the potential luminosities with the Eddington luminosity of
NGC~1068.  The Eddington luminosity is based on the observations of
Gallimore et al.\ (1996b) and Greenhill et al.\ (1996) who used water
maser emission lines to estimate the mass of the black hole in
NGC~1068. Unfortunately, the estimates of the two groups differ by a
factor of three.  Gallimore et al.'s measurements result in an
Eddington luminosity of 1$\times$10$^{12}$L$_\sun$ while those of
Greenhill et al. give 3$\times$10$^{11}$L$_\sun$. Both these
luminosities are less than the putative isotropic bolometric luminosity
of $\gtrsim$25$\times$10$^{11}$L$_\sun$ derived above, but we can
strongly discount this value only if we accept Greenhill et al.'s value
of the black hole mass. It should be noted that either value of the
Eddington luminosity is only  somewhat larger than the total
mid-infrared luminosity suggesting that the accretion onto the black
hole is almost Eddington-limited and that any gas within the ionization
cone will be accelerated outward by radiation pressure as the flux
there must exceed the  Eddington flux.

The radiation pressure should also sweep the ionization cone clean of
dust, although there is evidence (Miller et al.\ 1991) that the grains
are continuously being replenished. The time scale for such clearing
would be quite short so an evacuated cone with dust at the edges might
result. Thus, with the exception of the dust in a hypothesized central
torus, the mid-infrared emitting dust would be distributed in a thin
shell around the surface of the ionization cone, and thereby appear as
linear features since the line-of-sight path length is maximized at the
edges of the shell.  The observations would most likely imply a very
patchy shell. This would obviate the need to invoke any beaming of the
ultraviolet radiation which heats the dust, other than the relatively
broad ionization cone.

Although the morphological resemblance of the radio and mid-infrared
emitting regions might suggest the radio jet being the heating source
of the dust in the tongue, we argue that this is unlikely based on the
energetics.  Observation shows that the plasma responsible for the
scattering of the AGN light is moving outward from the nucleus at a
velocity of 300--1200 km/s (Miller et al.\ 1991).  If
we take this plasma velocity as the shock speed, mid-infrared emission
can be produced by shocks in this flow through two dust-heating
mechanisms:  direct collisions between the plasma and dust particles
(Draine 1981) and absorption of ultraviolet radiation produced by the
post-shock cooling plasma (cf. Dopita \& Sutherland 1996).  However,
unless the efficiency of converting the shock energy into infrared
emission is extremely high, we cannot avoid the conclusion that this
shock carries a mechanical luminosity comparable to the Eddington
limit, which seems unlikely.  There is no evidence that jets in Seyfert
galaxies are this luminous.  Furthermore, regardless of which mechanism
dominates the dust heating, such a fast shock is bound to produce a
substantial ultraviolet  radiation.  Therefore, the absence of strong
ultraviolet sources near the mid-infrared emitting region (recall that
the ultraviolet radiation in the nuclear region is mostly scattered AGN
light) also argues against the shock heating by the radio jet unless
the ultraviolet radiation is perfectly shielded by dust.

Our picture, then, is of a thick, radiation-dominated accretion disk
close to the central black hole that emits an ultraviolet power
somewhat larger than the total mid-infrared  luminosity which,
primarily through electron scattering, with perhaps some dust
absorption, is collimated into a pair of roughly anti-parallel cones.
It is likely that these cones are not just framed by a single accretion
torus  but are defined successively on many scales from the radius of
the black hole  to the radius of the narrow line region.  Indeed, the
partial occultation  of the ionization cone by the dense, dusty gas
that we are observing may be just a final stage in this process.

Why is the mid-infrared emission only found mainly on one side of the
ionization cone and the radio jet? We suspect that this is because the
gas flow  on all scales may be quite irregular and the beam of the
ionizing radiation is  far from axisymmetric.  Precession of the
central accretion disk may also contribute to this outcome. This can
also account for the otherwise  surprising observation that the color
temperature actually increases along the tongue. Suppose that the
intensity in the northern ionization cone is quite non-uniform due to
the irregular scattering and absorption associated with its formation.
The hotter dust observed near  the end of the tongue might then be
associated with an unusually  intense pencil impacting a dense and
dusty cloud, where the radiation  flux is larger than the  radiation
flux that heats dust closer to the core despite the inverse square law
dilution.  In this way, we believe that we can account qualitatively
for the observed morphology from  radio to ultraviolet wavelengths.

If the  interpretation presented above survives further scrutiny, it
suggests an explanation as to why Seyfert galaxies (and most quasars)
are radio-quiet (not silent).  Specifically  the radiation drag close
to the black hole and within the ionization cone may prevent  the
outflow from attaining ultra-relativistic speed. Future observations of
other nearby Seyfert galaxies will be necessary to see if total
luminosities are as large as found  for NGC~1068.

In order to substantiate this picture it would be useful to show that
the AGN has sufficient x-ray, ultraviolet, and visual luminosity to
provide  for the  observed mid-infrared luminosity. Observations of
visual polarization, like those of Capetti et al\ (1995), should be
able to give this since it is presumably due to dust scattering. If we
assume that the grains are similar to Galactic grains, their blue polarized flux should correlate with the mid-infrared flux. We can use the
albedo and polarization  expected from dust scattering  to estimate the
luminosity absorbed. This should give a lower limit on the luminosity in the ionization cone..

The luminosities measured in the present images are, as noted, close to
the Eddington luminosity for NGC~1068. If the tight correlation between
the bulge velocity dispersion and  black hole mass (Ferrarese \&
Merritt 2000; Gebhardt et al.\ 2000) is substantiated, black hole
masses may be inferred in other galaxies from velocity dispersion
measurements.  By comparing the corresponding Eddington luminosity with
the AGN luminosity derived from infrared observations similar to those
presented here, we will be able to determine which AGN are accreting at
the Eddington rate. This should enhance our understanding of how black
holes are fueled in galactic nuclei.

Finally, a consequence of the picture presented above is that a
significant fraction of the luminosity of NGC~1068 is not associated
with a torus \textit{per se}, but is produced by re-processing
radiation with dusty gas located $\sim$40~pc away from the central
source whose characteristics may be unique to NGC~1068. This complexity
makes any interpretation of AGNs invoking a common morphology more
difficult.

\section{Summary and Conclusions} 

1) Mid-infrared observations of NGC~1068 at the diffraction limit of
the Keck 10-m Telescope show a bright core containing about 1/3 the
total flux accompanied by north-south emission about 70~pc long which
is unresolved in the east west direction. Thus most of the emission
from the nuclear region of NGC~1068 comes from the elongated, highly
asymmetric  north-south structure.

2) At wavelengths between 7.9 and 24.5~$\mu$m, essentially all the 
nuclear flux of NGC~1068, as distinct from that in the 3~kpc
ring, is contained within a diameter of 300~pc of the
central core.

3) The 12.5 to 24.5~$\mu$m color temperature of the radiation is
$\sim$270~K and does not decrease smoothly with distance from the
core, but has a secondary maximum close to 270~K after dropping to
$\sim$215~K.

4) The absorption by silicates is lumpy even within the central
region of NGC~1068. There is evidence for increased silicate absorption
over  the central core and to the south; there is less silicate
extinction to the knots of emission to the north.

5) The low level mid-infrared emission, when located using the 
scattering center determined by Kishmoto\ (1999), lies near one 
wall of the conical
region depicted in [OIII]  (Evans et al.\ 1991; Macchetto et
al.\ 1994), and features in the near-infrared emission correlate
with clumps in the [OIII] emission. The mid-infrared emission
mimics the radio jet observed by Gallimore et al.\ (1996a). 

6) No evidence for the putative outer torus present at 3.5 and
4.8~$\mu$m  (Marco \& Alloin 2000) is seen in these mid-infrared
images.

Two thirds of the mid-infrared emission -- the core --is interpreted as
being the result of re-radiation of the AGN luminosity by a dusty torus
viewed at an angle so we see one heated face almost directly while the
other face is obscured. Most of the ultraviolet power of the nucleus
appears to be redirected into the ionization cone by electron
scattering.There are, however, significant details which need to be
addressed before this picture can be accepted. The remaining third of
the mid-infrared emission in NGC~1068 -- the tongue -- is associated
with the western boundary of both the [OIII] emission and the radio
continuum emission.  We identify this with re-emission  by dust of
visual and ultraviolet radiation which originates in the core of
NGC~1068 and is strongly beamed along the ionization cone.

The ability to perform ground-based imaging photometry of active
galaxies in the   mid infrared is opening up several scientific
opportunities. The most immediate is that observations such as these
will greatly improve our estimates of the cosmological luminosity
density of accreting  black holes. X-ray observations have demonstrated
that there is a large population of heavily obscured AGN. X-ray
observations however, are not able to measure the absorbed power with
any precision, and for this, infrared observations  are  crucial. We
suspect that most re-radiation by dust in obscured AGN emerges in the
mid infrared. By contrast, most emission from star forming regions may
occur at longer wavelengths. In order to test this hypothesis, it
will  be necessary to perform additional, high resolution observations
in the mid infrared of active galaxies, like those described here. If
the hypothesis is confirmed, then SIRTF will be able to measure the
re-radiation in a much larger sample and provide the first accurate
measurement of the luminosity density of active galaxies.

\newpage

\acknowledgments{}
We thank the staff of the Keck Observatory, especially Bob Goodrich,
for their assistance in making these observations possible. We also
thank Pat Shopbell for discussions about  the x-ray observations of
NGC~1068 and Lee Armus, Bob Goodrich, Nick Scoville, Kris Sellgren, and
Dave Thompson for discussions about various aspects of the
interpretation.  We thank Ski Antonnucci and Makoto Kisimoto for their
helpful comments as referees.  The W.~M.~Keck Observatory is operated
as a scientific partnership between the California Institute of
Technology, the University of California and the National Aeronautics
and Space Administration.  It was made possible by the generous
financial support of the W.~M.~Keck Foundation. B.T.S., G.N., K.M. and
E.E. are supported by grants from the NSF and  NASA. B.T.S. is
supported by the SIRTF Science Center at Caltech. SIRTF is carried out
at J.P.L., operated by Caltech under an agreement with NASA. This work
was carried out in part (J.J.B.,M.W.W. and M.E.R.) at J.P.L. The
development of MIRLIN was supported by NASA's Office of Space Science.

\appendix
\section{Appendix A -- Tests of Deconvolution}

There have been several, apparently conflicting, images of NGC~1068 in
the  infrared recently published. We therefore describe below in detail
the tests performed on our images to characterize  the statistical
uncertainties and stability of the PSF, and to uncover potential
sources of systematic uncertainty.  Because noise models are difficult
to estimate in practice for  deconvolved images, multiple redundant
observations were used to make an assessment  of the uncertainties.
For example, from a comparison of the raw and re-convolved images, the
differences between the images are found to be generally largest near
the brightness peak, at  a level significantly larger than the
estimated statistical noise.  This comparison implies that the
uncertainties in the deconvolved images are dominated by variations in
the shape of the  PSF during the observations, as might be expected
given the high signal-to-statistical-noise  ratio in the raw images.

Operationally, the uncertainty was characterized by comparing six
deconvolved images  based on six independent sets of NGC~1068 and
PSF raw images (derived from the image  pairs in Table~1).  These
deconvolved images (see Figure~7) consistently reproduce the
general features from the combined deconvolved image and the
uncertainty over the image can be evaluated by comparison of
these  deconvolved images.  As shown in  Figure~7, the
uncertainties are largest near the infrared peak, and decrease
with distance from the peak.   The uncertainty in the combined
image is $\sim$5\% of the surface brightness. 

As a check of the robustness of the deconvolved images, two  images of
NGC~1068 and PSF were constructed by combining data throughout the
observations.  The data  were interleaved such that the effective
airmass of the NGC~1068 and PSF were similar.   The deconvolved images
(NGC~1068 \#1 and \#2 in Figure~8) are in excellent qualitative
agreement with each other and the deconvolved images presented in
Figure~1.  The faint  source to the northeast is present in all of the
images at 10.3, 12.5 and 24.5~$\mu$m.  The largest  disagreement occurs
at 7.9~$\mu$m, where the second image shows a rotated central
elongation and a faint source, 1.4\% of the peak, about 0.2$''$
southwest of the nucleus. By so dividing , we see that The major
features (east-west width, shape of the northern and southern
extensions, relative strength of the  peak  brightness) are seen to be
reproducible when the data are divided into individual images.  Faint
sources, such as the northeast source, generally appear in  each of the
deconvolved images in Figure~7 and Figure~8.  One exception is the
faint  source southwest of the nucleus at 7.9~$\mu$m, which is possibly
spurious.

As a test of systematic uncertainty, PSFs were  constructed from
HR~1457 and HR~0911.   The PSFs were mismatched both in average
airmass ($\sim$1.15 for HR~0911 and 1.03 for  HR~1457) and in flux
(HR~1457 is 2.8 times brighter than HR~0911).  At 12.5~$\mu$m it is
also possible to deconvolve the October~1 data using the
October~3 PSF observations. The deconvolution of the one PSF by
the other is shown in Figure~8. We chose to deconvolve the PSF
with larger average airmass by the PSF with the lower average
airmass to illustrate the maximum error. Since  the FWHM of the
PSF observed at a lower airmass is always slightly smaller than
that of the PSF with a higher airmass, the reverse deconvolution
gives significantly better results.  The deconvolved PSFs are
ellipsoidal, with FWHM in the extended direction that ranges from
1.5 to 2 times that of the deconvolved PSFs presented in Figure~1.
The largest spurious features in Figure~8  appear at 3\% of the
peak brightness, due to incomplete cancellation of the first Airy
diffraction ring.

As a further test, images of NGC~1068 were deconvolved with a
mismatched  calibrator.  As shown in Figure~8,  a combined image of
NGC~1068 observed  in sequence with HR~1457, and deconvolved using the
PSF from HR~0911 (NGC~1068 \#3) was formed.   The remaining
observations of NGC~1068 were deconvolved using the PSF from HR~1457
(NGC~1068 \#4).  The two deconvolved images gave similar results,
although the NGC~1068 image and the PSF were mismatched in airmass as
much as possible, and well separated in time.  Finally, as a severe
test of the systematic uncertainty, the image of  NGC~1068  from the
night of 3~October was deconvolved by the PSF from 1~October (NGC~1068
\#5) and the 1~October  observations of NGC~1068 were deconvolved by
the PSF from 3~October (NGC~1068 \#6).  Although these  results are
less satisfactory, as might be expected, the general morphology is
clearly  preserved. Both these deconvolved images of the NGC~1068 and
the PSF illustrate  the maximum effect of systematic error due to
changes in the PSF with time and elevation angle.  Both the
differential average airmass and time separation between raw image and
PSF are many times better for the images presented in Figure~1.

As a final test, images of NGC~1068 and a PSF star (IRC~+00~032) were
obtained on 27~January, 2000 using a  different mid-infrared camera,
the LWS (Jones \& Puetter 1993) on the Keck~1~10-m Telescope.   LWS
provides Nyquist-sampling of the Airy function with 0.08$''$ pixels.
Unfortunately, on the night of the observations  the seeing was
approximately twice the diffraction-limited resolution. One hundred and
ninety two observations, in six groups of 32 observations and each of 1
sec exposure time, were taken of NGC~1068 interleaved with similar
observations of the PSF star.  The 10\% of each set of $\sim$~200
observations with the smallest FWHM were selected and the images
coadded; these coadded images were used to obtain  a deconvolved image
of NGC~1068 (Figure~8) that bears a good resemblance to the images
obtained  with MIRLIN.

\section{Appendix B -- Tests of Sensitivity to Additional Sources}

The detectability and stability of a source near the main peak in the
deconvolved image may be tested by creating a synthesized image,
whereby a faint point source is added to the raw NGC~1068 image.  If
the PSF is systematically different in the PSF and NGC~1068 images, the
deconvolved synthesized image may be expected to change the flux
density and position of the faint point source.  A point source based
on the deconvolving PSF with a flux density ranging from 10\% - 0.5\%
of the flux density of NGC~1068 was placed at positions $\sim$0.5$''$
east and west of the nucleus.  The point source was generally recovered
in the deconvolved image with relatively little change in flux density
or position down to the 0.5\% level at 7.9, 10.3, and 12.5~$\mu$m, and
down to the 2.0\% level at 24.5~$\mu$m.  These levels are consistent
with the contours chosen in Figure 1.

If the point source is inserted closer to the nucleus (0.3$''$ to the
east or west),  the flux density and position are still recovered in
the deconvolved image.  The deconvolved resolution appears to slightly
degrade so that peak surface brightness is about two times lower.
These tests were also performed using an independent PSF (HR~0334) in
the synthesized image.  In this case, the resolution of the point
source was somewhat lower, as expected.  However, the flux density and
position are recovered with similar accuracy as before. Based on these
tests, the estimated uncertainty in the photometry of a source placed
0.3$''$ -- 0.6$''$ east or west of the nucleus, evaluated in a
0.3$''\times$0.3$''$ artificial beam, is $<$0.5\% of the total flux
density at 7.9 and 12.5~$\mu$m.

Similar tests were also performed on the LWS images described in
Appendix A with  worse  image recovery than with the MIRLIN data.  The
seeing conditions when  the LWS data were taken were significantly
poorer than when  the MIRLIN data were obtained and limited the
reliability of the LWS data.

\clearpage

\figcaption 
{a--Four images from the present data are given for wavelengths
7.9~$\mu$m (top row), 10.3~$\mu$m (second row); 12.5~$\mu$m (third
row); 24.5~$\mu$m (bottom row). The first two columns show the raw
images of the appropriate PSF and of NGC~1068. The right hand  two
columns show the deconvolved images of the appropriate PSF and of
NGC~1068.  Note that the size scales are different between the raw and
deconvolved images. North is up and east is to the left in all the
images. The maximum contour shown is of pixels with a flux density 
surface brightness of 0.9$\times$[maximum flux density surface 
brightness per pixel]. The minimum contour shown is down by a 
factor of 140 from the peak surface brightness in the 7.9, 10.3, 
and 12.5~$\mu$m images, and by a factor of 70 in the 24.5~$\mu$m 
image.  Contour levels are spaced by multiplicative factors of two.\\
\hspace*{.7in}b--The deconvolved image at 12.5~$\mu$m,  which has the
highest spatial resolution and the highest signal to noise ratio, i.e.,
the highest dynamic range, is duplicated at a larger scale than in
Figure~1a. The contour spacing in the NGC~1068 image is $\sqrt{2}$;
the other parameters of the contour levels are as in Figure~1b. 
The contour for the PSF is drawn at half of the maximum flux 
pixel$^{-1}$. The dashed horizontal line marks the arbitrary 
demarcation between the core --to the south -- and the tongue -- 
to the north.}

\figcaption  
{The spectral energy distribution of NGC~1068 within a 4$''$ diameter
beam resulting from the present observations (open circles; dashed
line) is presented along with representative measurements from the
published literature (Neugebauer et al.\ 1971; Marco \& Alloin,\ 2000 ;
Penston et al.\ 1974; Telesco et al.\ 1984; Thronson et al.\ 1987;
Benford 1999; Joint IRAS Science Team 1989).  The latter compilation is
not meant to be exhaustive and represents various sizes of big beam
measurements and includes the emission from the ``3~kpc ring'' (Telesco
et al.).  The arrows represent conservative upper limits on the
flux in a point source located at the core of NGC~1068 in the
deconvolved images.}

\figcaption 
{A contour plot of the 12.5~$\mu$m deconvolved image is overlaid on
the O[III] image of Machetto et al.\ (1994).  The contours go from 0.9
of maximum in multiplicative steps of a factor of two as in Figure 1a.
No absolute astrometry was obtained for the Keck images; the
juxtaposition of the infrared core with the [OIII] image was obtained
by aligning the mid-infrared peak with the position of the nucleus
determined by ultraviolet  polarimetry (Kishimoto\ 1999) shown by the blue
cross.  The letters label the local maxima in the [OIII] emission as
described by Evans et al.\ (1991).  The somewhat subjective boundary
of the conical region, denoted by dashed lines, is drawn to include
[OIII] clouds in the southern cone with strong polarization.
Determinations of the position of the nucleus
at near-infrared wavelengths (Thatte et al.\ 1997) and mid-infrared
wavelengths (Braatz et al.\ 1993) are shown by the green square and
circle, respectively.  These independent determinations of the position
of the nucleus agree within the quoted uncertainties, designated by 
the size of the symbols.} 

\figcaption 
{A contour plot of the 12.5~$\mu$m deconvolved image is superimposed
on the 5 GHz map of Gallimore et al.\ (1996c). The contours are chosen
as in Figures~1a and 3.  The mid-infrared peak is aligned to the S1 radio
source; no absolute astrometry was obtained for the Keck images.
The position of source S1 coincides with the ultraviolet polarization 
center within the astrometric uncertainty (Kishimoto\ 1999, 
Capetti et al.\ 1997).}

\figcaption
{The spectral energy distribution in artificial beams of 0.2$''$
diameter at five places in the 12.5~$\mu$m deconvolved image is given
in the left hand panel.  The locations of the photometry are shown in
the right hand panel superimposed on a contour plot of the 12.5~$\mu$m
deconvolved image smoothed to 0.2$''$ resolution.  The contour levels
are spaced by a multiplicative factor of two from the maximum; the size
of the artificial beam is indicated.  Each of the four deconvolved
images of Figure~1 were smoothed to similar resolution to make this
comparison.  Silicate absorption is identified with the  10.3~$\mu$m
data.}

\figcaption 
{Color temperatures derived from the 12.5 and 24.5~$\mu$m deconvolved
images are shown.  The deconvolved images were first smoothed to
0.2$''$ to insure comparable resolution in both images.  The minimum
flux density surface brightnesses used in obtaining the color
temperatures were 0.03 $\times$~the maximum flux density pixel$^{-1}$
at 12.5~$\mu$m and 0.12 $\times$~the maximum flux density pixel$^{-1}$
at 24.5~$\mu$m.The contours go from a maximum of 269~K to a minimum of
169~K in intervals of 20~K. The dashed contours represent  the
12.5~$\mu$m deconvolved image smoothed to 0.2$''$ resolution.  The
contour levels are spaced by a multiplicative factor of two from the
maximum.
 }

\figcaption
{Deconvolved images based on pairs of raw NGC~1068 and PSF images at
12.5~$\mu$m are shown.    The labels correspond to the numbers of the
image pairs in Table~1, with the bright star used as a PSF and the
air masses are identified therein.  Image \#7 has the poorest matching
between the airmass of NGC~1068 and that of the PSF. Contours are
spaced by  multiplicative factors of 2 as in Figure~1.  The combined
image and estimated uncertainties are shown at the right with the same
minimum contour.}

\figcaption
{A comparison of deconvolved images derived from subsets of the
observation log listed in Table~1 is given. The columns in the upper
left 3$\times$4 panels are:  (NGC~1068~Total) the  deconvolved image of
NGC~1068 using the entire data set, (NGC~1068~\#1 and NGC~1068~\#2)
deconvolved images of NGC~1068 using half of the NGC~1068 and PSF
data.  The data  were interleaved to have similar average air masses
for the NGC~1068 and the PSF observations.   In  the upper right
1$\times$4 panel, from top to bottom are:  the deconvolved image of
NGC~1068 obtained with the LWS instrument at a wavelength
$\lambda$~$\approx~$7.9~$\mu$m under non-ideal seeing conditions,
(PSF~\#3) the deconvolved PSF at $\lambda$~=~12.5~$\mu$m from the
October~1 PSF deconvolved by  the  October~3 PSF, and (PSF~\#2) the
deconvolved PSF from the HR~0911 PSF deconvolved by the HR~1457 PSF at
$\lambda$= 12.5 and 24.5~$\mu$m.  In the bottom 4$\times$1 panel:
(NGC~1068~\#3) the image of NGC~1068 obtained when HR~0911 was
substituted as a PSF instead of HR~1457 using a combined image of
NGC~1068 when HR~1457 was the PSF pair in Table~1, (NGC~1068~\#4) the
image of NGC~1068 obtained when HR~1457 was substituted as a PSF
instead of HR~0911 using a combined image of NGC~1068 when HR~0911 was
the PSF pair in Table~1 , (NGC~1068~\#5) the NGC~1068  deconvolution
combining the NGC~1068 observations of October~3 and PSF observations
from October~1, (NGC~1068~\#6) the NGC~1068 deconvolution combining the
NGC~1068 observation from October~1 with the PSF observation from
October~3.
}

\clearpage

\normalsize
\singlespace

\newpage
\begin{deluxetable}{cccccccc}
\tablewidth{0pt}
\tablenum{1}
\tablecaption{Log of Observations}
\tablehead{
   \colhead{Object}&\colhead{Image}&\colhead{7.9}&\colhead{10.3}&\colhead{12.5}&\colhead{24.5}&\colhead{U.T.}&\colhead{airmass}\\
\colhead{}&\colhead{No.}&\colhead{$\mu$m}&\colhead{$\mu$m}&\colhead{$\mu$m}&\colhead{$\mu$m}&\colhead{h:m}&\colhead{}\\
}
\startdata
\multicolumn{8}{c}{1 October 1998} \\
  NGC~1068   & 1 &         &         & $\surd$ &         & 10:09 & 1.27 \\
  HR~0911 & 1 & $\surd$ &         & $\surd$ & $\surd$ & 10:21 & 1.26 \\
  HR~0911 & 2 & $\surd$ &         & $\surd$ & $\surd$ & 10:54 & 1.16 \\
  NGC~1068   & 2 & $\surd$ &         & $\surd$ & $\surd$ & 11:17 & 1.11 \\
  HR~0911 & 3 & $\surd$ &         & $\surd$ & $\surd$ & 11:43 & 1.08 \\
  NGC~1068   & 3 & $\surd$ &         & $\surd$ & $\surd$ & 12:14 & 1.06 \\
  HR~1457 & 4 & $\surd$ &         & $\surd$ & $\surd$ & 12:46 & 1.08 \\
  NGC~1068   & 4 & $\surd$ &         & $\surd$ & $\surd$ & 13:15 & 1.09 \\
  HR~1457 & 5 & $\surd$ &         & $\surd$ & $\surd$ & 13:33 & 1.01 \\
\multicolumn{8}{c}{3 October 1998} \\
  HR~0911 & 6 &         & $\surd$ & $\surd$ &         & 12:32 & 1.03 \\
  NGC~1068   & 6 &         & $\surd$ & $\surd$ &         & 12:50 & 1.08 \\
  HR~0911 & 7 &         &         & $\surd$ &         & 13:18 & 1.05 \\
  NGC~1068   & 7 &         & $\surd$ & $\surd$ &         & 14:13 & 1.22 \\
\enddata
\end{deluxetable}
\newpage

\begin{deluxetable}{cccc}
\tablewidth{0pt}
\tablenum{2}
\tablecaption{Photometry for NGC~1068}
\tablehead{

\colhead{Wavelength}&\colhead{Magnitude\tablenotemark{a}}& \colhead{f$_\nu$\tablenotemark{a,b}}&\colhead{Mag Diff\tablenotemark{c}} \\
\colhead{$\mu$m}&\colhead{mag}&\colhead{Jy}&\colhead{mag}
}
\startdata
 7.9 &  1.19 & 21$\pm$2 & 0.04 \\
10.3 &  0.46 & 25$\pm$2 & 0.06 \\ 
12.5 & -0.42 & 38$\pm$4 & 0.08 \\
24.5 & -2.58 & 75$\pm$8 & 0.16 \\ 

\enddata
\tablenotetext{a}{4$''$ diameter beam}
\tablenotetext{b}{f$_\nu$ for 0.0 mag based on IRAS formulation 
(Beichman et al.\ 1985)}
\tablenotetext{c}{Magnitude difference between photometry in a 
4$''$ and 2$''$ diameter beam}

\end{deluxetable}

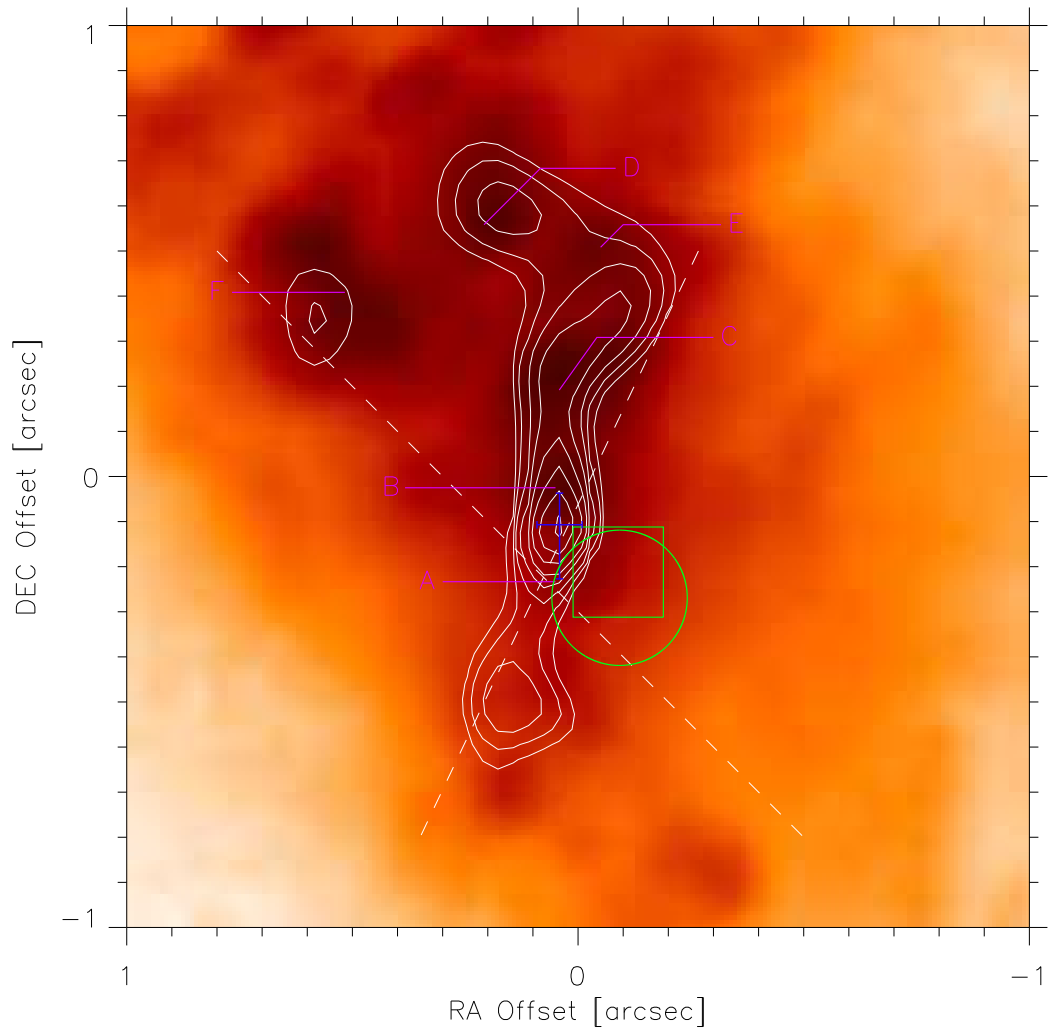
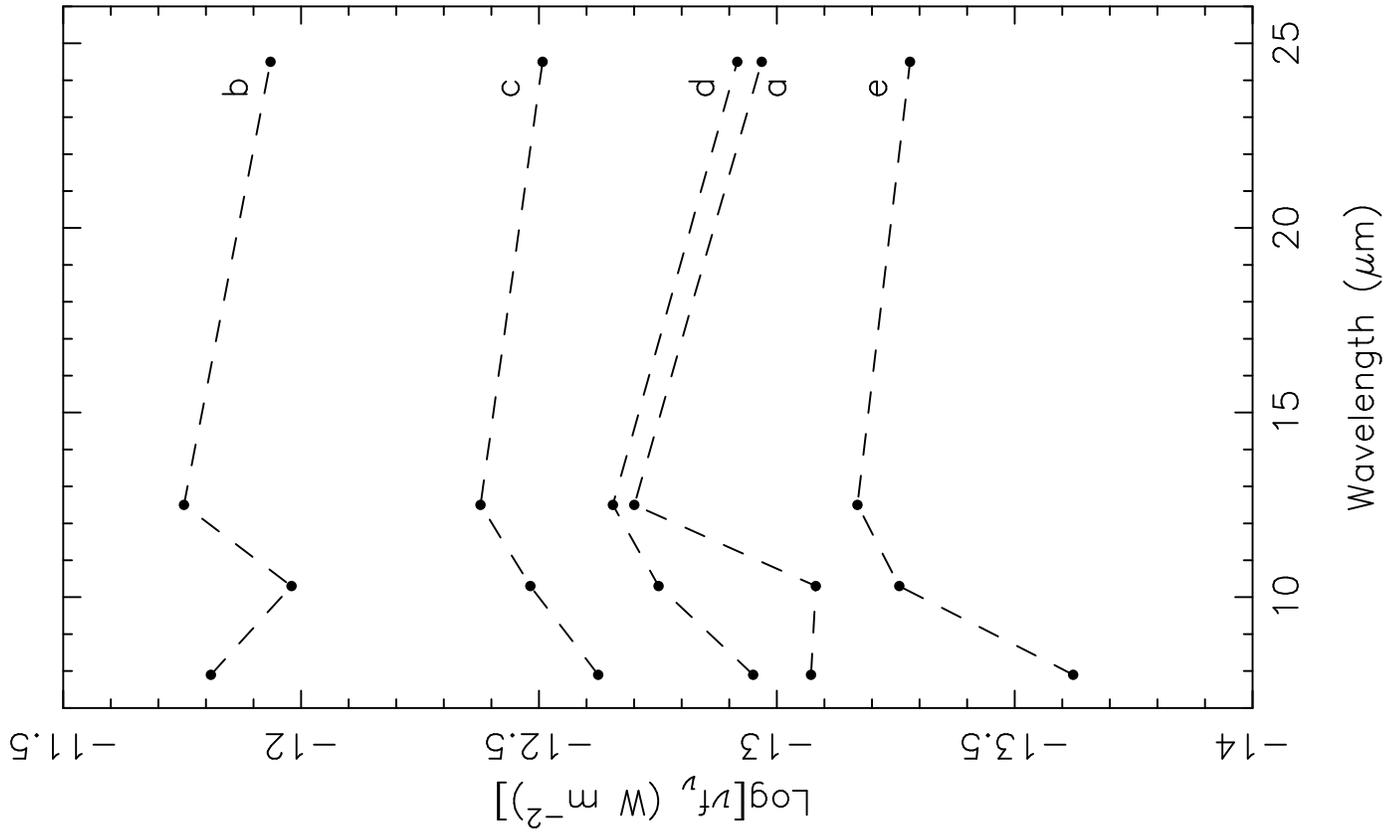
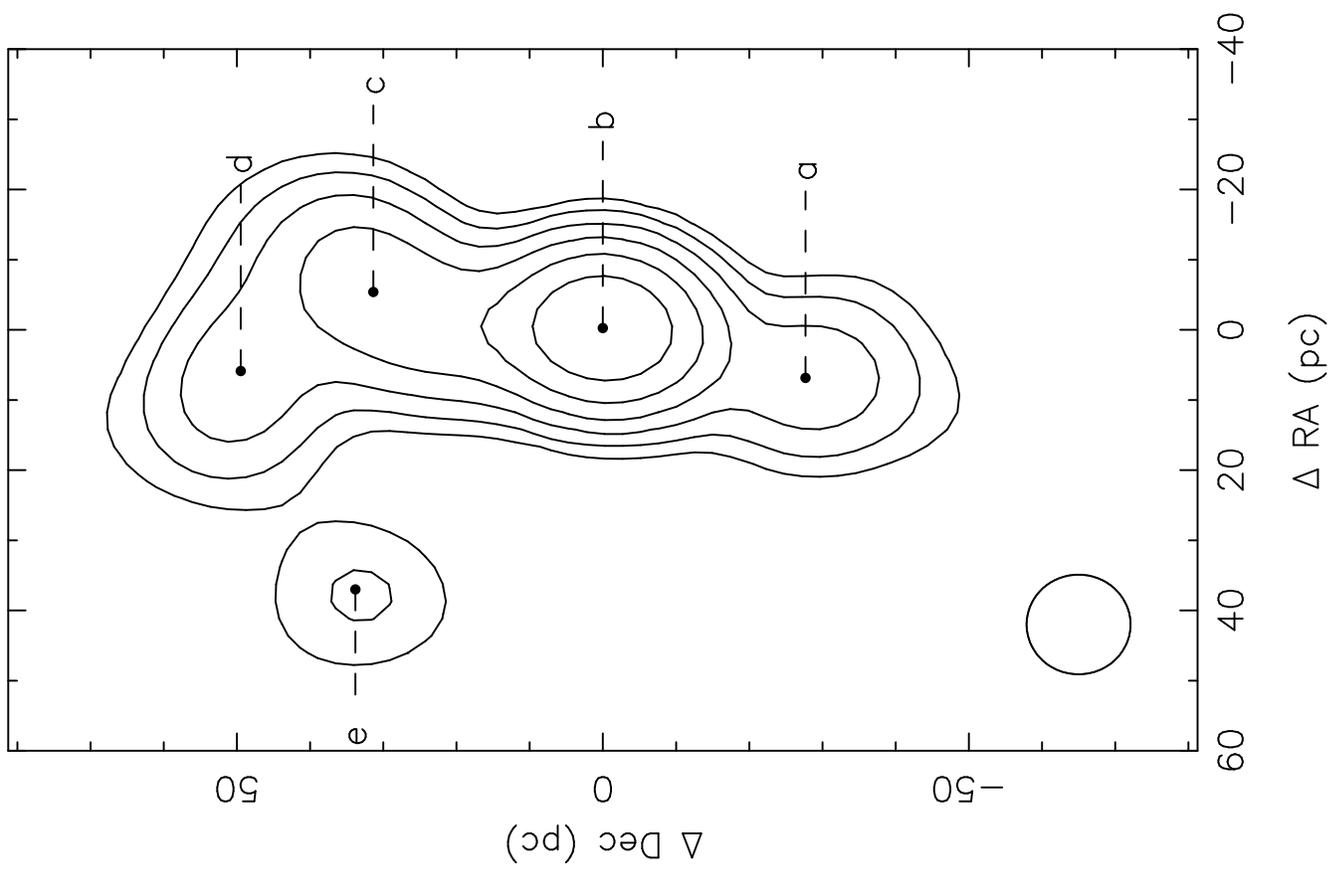
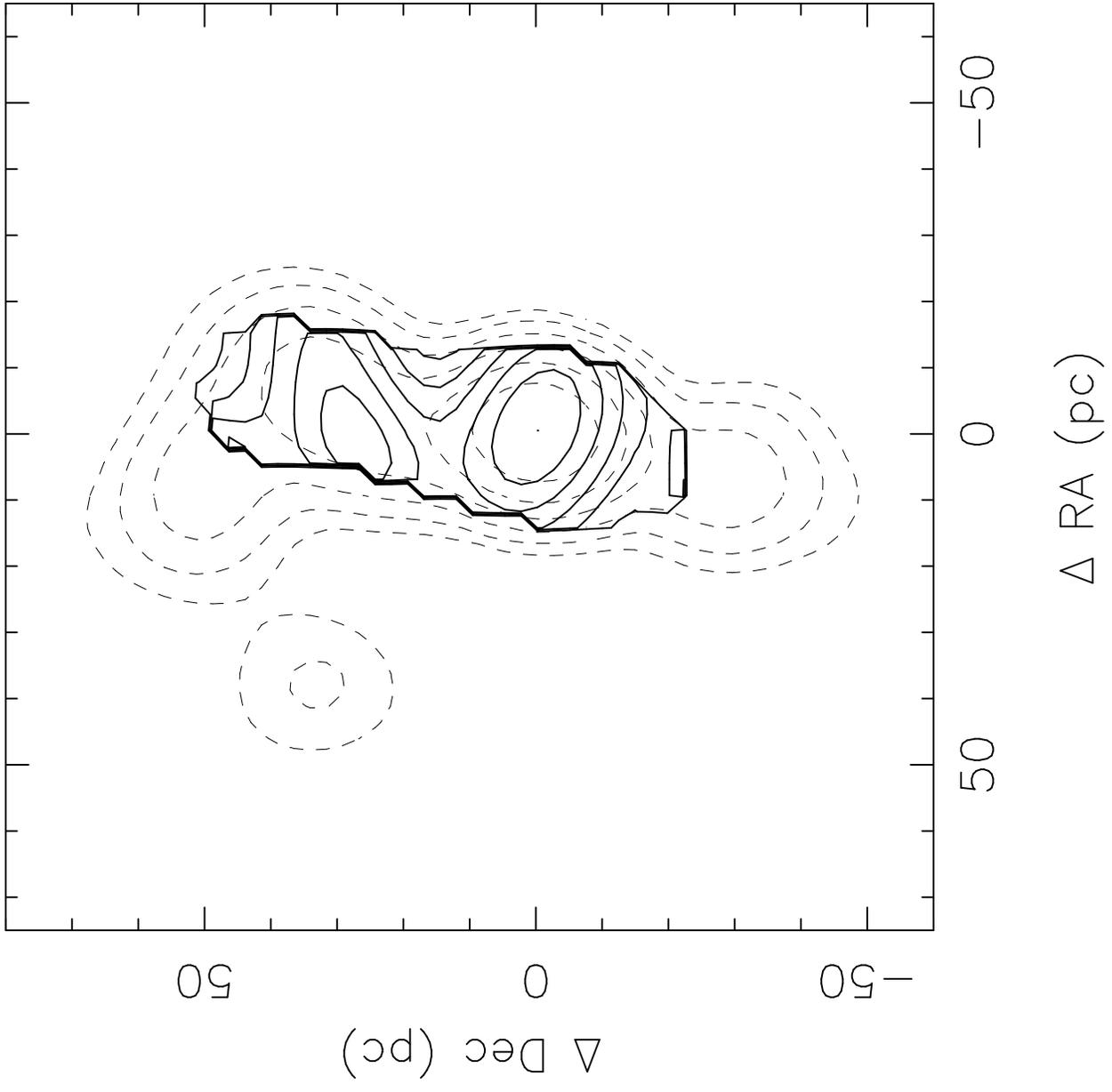
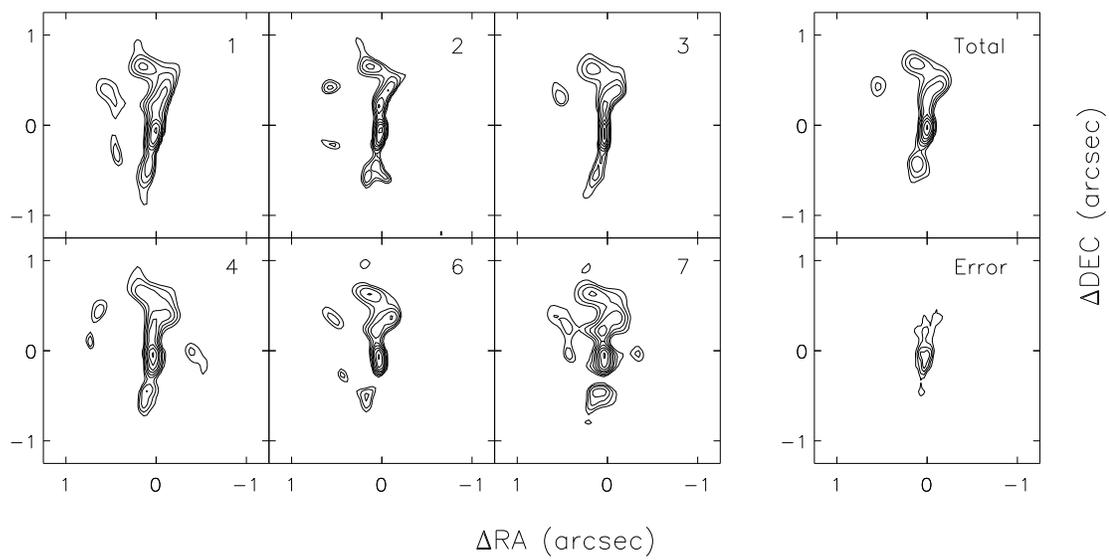
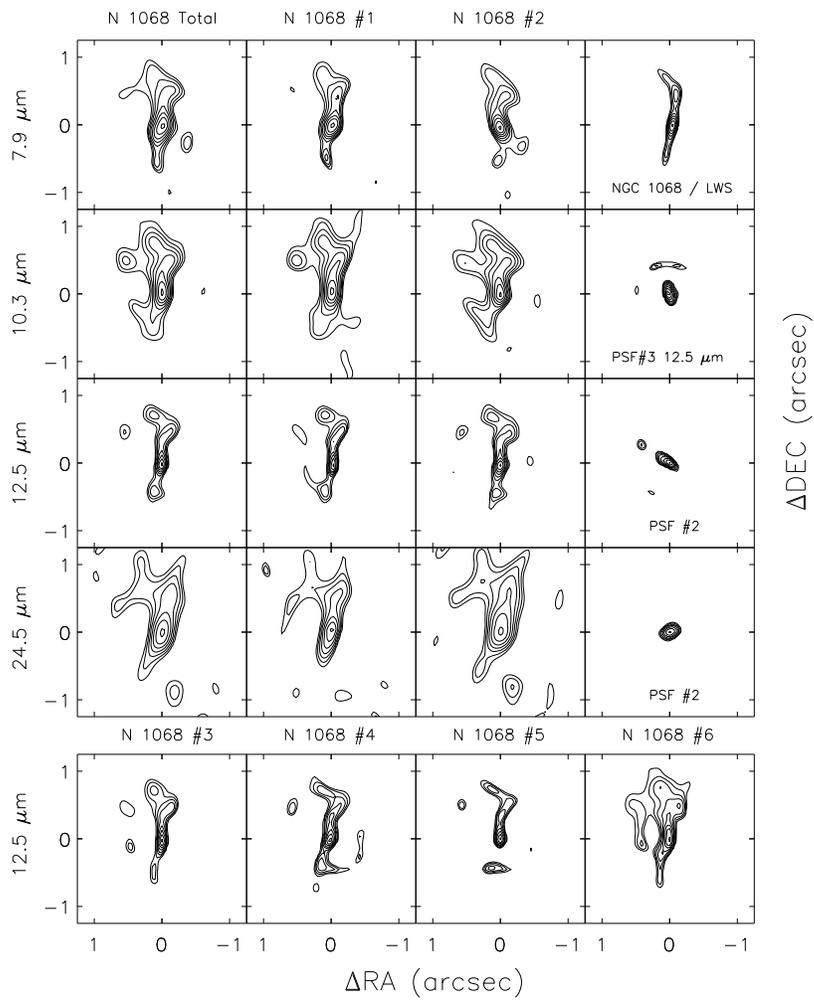
\end{document}